\documentclass[12pt,a4paper]{elsart}

\usepackage{epsfig}
\usepackage{amsmath}
\usepackage{graphics}
\usepackage{amssymb}

\begin{document}

\begin{frontmatter}



\title{Entropy Optimization of Scale-Free
Networks Robustness to Random Failures}


\author[Bing Wang]{Bing Wang} \ead{bingbignmath@yahoo.com.cn }
\author[Bing Wang] {Huanwen Tang\corauthref{cor}}
\corauth[cor]{Corresponding author.} \ead{imsetang@dlut.edu.cn}
\author[Bing Wang]{Chonghui Guo}\ead{guochonghui@tsinghua.org.cn}
\author{Zhilong Xiu}\ead{zhlxiu@dlut.edu.cn}
\address[Bing Wang]{Department of Applied Mathematics, Dalian University of Technology, Dalian 116024, P. R. China}
\address[Zhilong Xiu]{School of Environmental and Biological Science and Technology, Dalian University of
Technology, Dalian 116024, P. R. China }
\begin{abstract}
Many networks are characterized by highly heterogeneous
distributions of links, which are called scale-free networks and
the degree distributions follow $p(k)\sim ck^{-\alpha}$. We study
the robustness of scale-free networks to random failures from the
character of their heterogeneity. Entropy of the degree
distribution can be an average measure of a network's
heterogeneity. Optimization of scale-free network robustness to
random failures with average connectivity constant is equivalent
to maximize the entropy of the degree distribution. By examining
the relationship of the entropy of the degree distribution,
scaling exponent and the minimal connectivity, we get the optimal
design of scale-free network to random failures. We conclude that
the entropy of the degree distribution is an effective measure of
network's resilience to random failures.
\end{abstract}

\end{frontmatter}

\section{Introduction}
Many complex systems are characterized by the network of
interactions among its components. It has been shown that most of
the networks are highly heterogeneous in their connectivity
patterns. Heterogeneity can be easily identified by looking at the
degree distribution $p(k)$, which gives the probability of having
a node with $k$ links. Most complex networks can be described by
degree distributions $p(k)\sim ck^{-\alpha}$, with
$\alpha\in(2,3)$. These networks include social networks(such as
movie-actor networks, science citations and cooperation networks),
the Internet and the World Wide Web, metabolic networks and
protein interaction networks, etc
\cite{Albert02,Newman03,Dorogovtsev02,Jeong00,Jeong01}.

Since Albert \emph{et al.} raised the question of random failures
and intentional attack on networks \cite{Albert00}, enormous
interest has been devoted to the study of the resilience of
networks to failures of nodes or to intentional attacks
\cite{Cohen00,Cohen01,Albert01,Holme02,Gallos04,Crucitti04}.
Magoni studied a general strategy of attacks on the Internet
\cite{Magoni03}. It is important to understand how to design
networks which are optimally robust against both failures and
attacks.

Most of researchers use percolation theory to study this problem
\cite{Cohen00,Zhou05}. A fraction $p$ of the nodes and their
connections are removed randomly$-$the integrity might be
compromised: for $\alpha>3$, there exists a certain threshold
value $p_{c}$. When $p>p_{c}$, the network disintegrates into
smaller, disconnected parts. Below that critical threshold value,
there still exists a connected cluster. For $2<\alpha<3$ , the
networks are more resilient and the threshold value $p_{c}$
approaches 1 \cite{Cohen00}. Also, some researchers work on
optimizing network robustness to both random failures and attacks
with percolation theory \cite{Valente04,Tanizawa05,Paul04}.

A simple but essential character of scale-free network is its
heterogeneous link distribution. Moreover, heterogeneity is in
direct relationship with the network's resilience to attacks. Many
real-world networks are scale-free and they are robust to random
failures but vulnerable to targeted attacks. Heterogeneity can be
measured by entropy \cite{Cancho03,Sole}. Sol\'{e} \emph{et al.}
used entropy of the remaining degree and mutual information to
study some model networks of different heterogeneity and
randomness \cite{Sole}.

In this paper, different from percolation theory, we try an
alternative point of view$-$the entropy of the degree distribution
to describe scale-free network's heterogeneity. To determine the
optimal design of scale-free network to random failures, we
maximize the robustness of scale-free network to random failures
while keeping the cost constant (that is, the average number of
links per node remains constant). We find that optimization of
scale-free network's robustness to random failures is equivalent
to maximize the entropy of the degree distribution. By optimizing
the entropy of the degree distribution, we get the scenario of
optimal design of scale-free network to random failures.
\section{Entropy of the degree distribution}
In general, scale-free networks have the degree distributions
$p(k)$, $p(k)\sim ck^{-\alpha}$, $k=m,m+1,\ldots,K$, where $m$ is
the minimal connectivity and $K$ is an effective connectivity
cutoff present in finite networks. Then the entropy of the degree
distribution can be defined as follows:
\begin{center}
\begin{equation}
H=-\sum_{k=1}^{N-1}{p(k)\log{p(k)}}
\end{equation}
 \end{center}
where $N$ is the total number of nodes in the network and the
restriction to the degree distribution is $p(k)\sim ck^{-\alpha}$.
Within the context of complex networks, the entropy of the degree
distribution provides an average measure of network's
heterogeneity, since it measures the diversity of the link
distribution \cite{Sole}. Two extreme cases are the maximal value
and the minimal one. The maximum value is $H_{max}=\log{(N-1)}$
obtained for $p(k)={\frac{1}{N-1}} (\forall k=1,2,\ldots,N-1)$.
$H_{min}=0$ occurs when $p(k)=\{0,\ldots,1,\ldots,0\}$.

For the power-law degree distribution of scale-free network, the
impact of diversity (long tails) is obvious, increasing the
uncertainty. As the scaling exponent increases or the cut-off
decreases, the network becomes less heterogeneous and as a result
a lower entropy is observed \cite{Sole}.

In this paper, we study the entropy of the degree distribution for
scale-free networks. With continuous approximation, for the power
law degree distribution of scale-free network, entropy $H$ can be
expressed as follows:
\begin{center}
\begin{eqnarray}
H&=&-\int_{m}^{K}{p(k)\log{p(k)}}dk\nonumber\\
&=&-\int_{m}^{K}{ck^{-\alpha}\log{(ck^{-\alpha})}dk}\nonumber\\
&=&-\frac{c\log{c}}{1-\alpha}\cdot{(K^{1-\alpha}-m^{1-\alpha})}+\frac{c\alpha}{1-\alpha}
  \cdot[\log{K}\cdot{K^{1-\alpha}}-\log{m}\cdot{m^{(1-\alpha)}}\nonumber\\
&&-\frac{1}{1-\alpha}\cdot{(K^{1-\alpha}-m^{1-\alpha})}]
\end{eqnarray}
 \end{center}
The largest connectivity $K\approx{mN^{\frac{1}{\alpha-1}}}$
\cite{Cohen00}, The coefficient $c$ is obtained from
\begin{center}
\begin{eqnarray}
1&=&\int_{m}^{\infty}{p(k)dk}=\int_{m}^{\infty}{ck^{-\alpha}dk}=\frac{c}{1-\alpha}\cdot{(-m^{1-\alpha})}\nonumber\\
\end{eqnarray}
 \end{center}
and thus
\begin{center}
\begin{eqnarray}
c=(\alpha-1)\cdot{m^{\alpha-1}}
\end{eqnarray}
 \end{center}
Then entropy $H$ can be expressed as a function of scaling
exponent $\alpha$, the minimal connectivity $m$ and network scale
$N$.
\begin{center}
\begin{equation}
H(\alpha,m,N)=(\log{(\frac{\alpha-1}{m})}+\frac{\alpha}{1-\alpha})\cdot{(\frac{1}{N}-1)}-\frac{\alpha}{\alpha-1}\cdot{\frac{\log{N}}{N}}
\end{equation}
 \end{center}

Fig. 1 displays the relationship between entropy $H$ and $\alpha$
for different $N$ when the minimal connectivity $m$ is defined.
\begin{figure}[th]
\centerline{\psfig{file=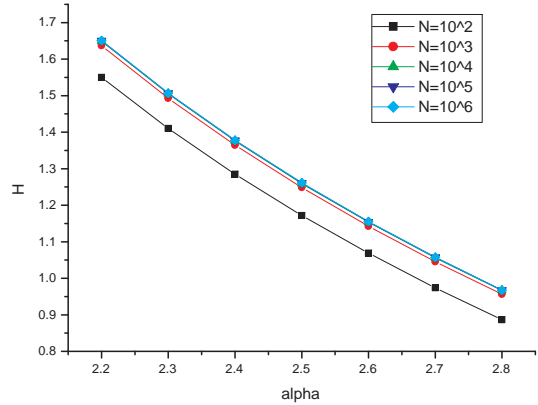,width=7cm}} \vspace*{8pt}
\caption{$H$ for different $N$ versus $\alpha$ with $m=1$.}
\end{figure}
We know that with the increase of scaling exponent $\alpha$, the
cut-off connectivity $K$ decreases, which makes the network less
heterogeneous. As a result, a lower entropy is observed. So
different scaling exponent $\alpha$ brings the network different
degree of heterogeneity.

The average connectivity $<k>$ is obtained with continuous
approximation,
\begin{center}
\begin{eqnarray}
<k>&=&\int_{m}^{K}{k\cdot{p(k)}dk}=\int_{m}^{K}{k\cdot{ck^{-\alpha}}dk}=\frac{c}{2-\alpha}\cdot{(K^{2-\alpha}-m^{2-\alpha})}
\end{eqnarray}
 \end{center}

and thus
\begin{center}
\begin{equation}
<k>\approx{\frac{\alpha-1}{2-\alpha}\cdot{m}\cdot{(N^{\frac{2-\alpha}{\alpha-1}}-1)}}
\end{equation}
 \end{center}

According to the above results, we can formulate the scale-free
network's resilience to random failures problem as the following
nonlinear mixed integer programming.
\begin{center}
\begin{eqnarray}
max&& H(\alpha,m,N)\nonumber\\
s.t.&&<k>=const \nonumber\\
\end{eqnarray}
 \end{center}
The problem mentioned above is different from the usual maximizing
the entropy of the network problem. Note that the highest entropy
for a network with a given $<k>$ is exponential random graph
\cite{Park04}. To get the optimal design of scale-free networks,
we restrict the degree distribution to be a power-law
distribution. The problem can be rewritten as follows:
\begin{center}
\begin{eqnarray}
max&&H(\alpha,m,N)\nonumber\\
s.t.&&\frac{\alpha-1}{2-\alpha}\cdot{m}\cdot{(N^{\frac{2-\alpha}{\alpha-1}}-1)}=const\nonumber\\
&&2<\alpha<3\nonumber\\
&&1\leq{m}\leq{(N-1)}\nonumber\\
&&m\in{integer}
\end{eqnarray}
 \end{center}
By solving the mixed integer programming problem, we get the
optimal value $H$, the optimal solutions $\alpha$ and
 $m$ for different average connectivity $<k>$. The results are shown in
Fig. 2.
\begin{figure}[th]
\centerline{\psfig{file=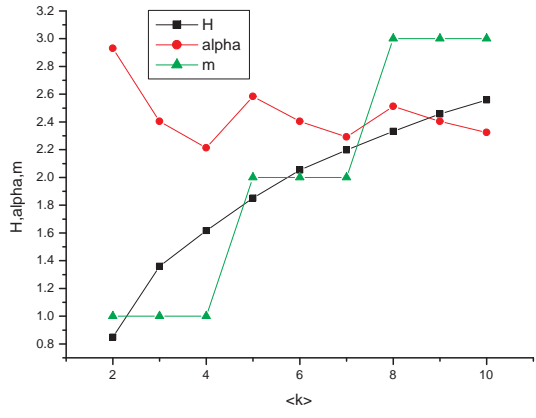,width=7cm}}
\vspace*{8pt} \caption{The optimal value $H$, the optimal
solutions $\alpha$ and $m$ versus $<k>$ with $N=1000$.}
\end{figure}

From Fig. 2, we can see that, in general, with the increase of
average connectivity $<k>$, entropy $H$ as well as the minimal
connectivity $m$ increases, too. In part, the optimal minimal
connectivity $m$ keeps constant for a certain range of $<k>$,
where scaling exponent $\alpha$ decreases with the increase of
$<k>$. As a result, the optimal network becomes much more
heterogeneous than the one with bigger scaling exponent.

To prove the entropy of the degree distribution is indeed an
effective measure of network robustness under node removal, we
examine the relationship between the entropy of the degree
distribution and the network threshold value $p_{c}$, which was
presented by Cohen \emph{et al.} with percolation theory
\cite{Cohen00}. The criterion to calculate the percolation
critical threshold value $p_{c}$ of a randomly connected network
is:
\begin{center}
 \begin{equation}
 1-p_{c}=\frac{1}{\kappa_{0}-1}
 \end{equation}
 \end{center}
where $\kappa_{0}=\frac{<k^{2}>}{<k>}$  is calculated from the
original distribution before the random failures. With
$K\approx{mN^{{\frac{1}{\alpha-1}}}}$, $\kappa_{0}$ can be
expressed as follows:
\begin{center}
\begin{equation}
\centering
\kappa_{0}\thickapprox\frac{\alpha-2}{3-\alpha}\cdot{m}\cdot{(N^{\frac{3-\alpha}{\alpha-1}}-1)}
\end{equation}
\end{center}

\begin{figure}[th]
\centerline{\psfig{file=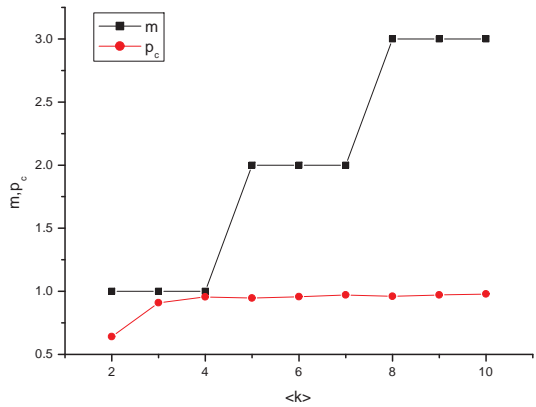,width=7cm}}
\vspace*{8pt} \caption{ $p_{c}$ and $m$ versus $<k>$ with
$N=1000$.}
\end{figure}
In Fig. 3, $p_{c}$ is calculated from the optimal value $\alpha$
and $m$ which are obtained from maximizing $H$ for different
average connectivity $<k>$. The results indicate that for some
constant $m$, $p_{c}$ increases with the increase of $<k>$.

Since both $\alpha$ and $N$ affect network's resilience to random
failures, we may ask how the network's robustness changes with
$N$. We just examine the relationship between $p_{c}$ and $\alpha$
for different $N$. The results are shown in Fig. 4.
\begin{figure}[th]
\centerline{\psfig{file=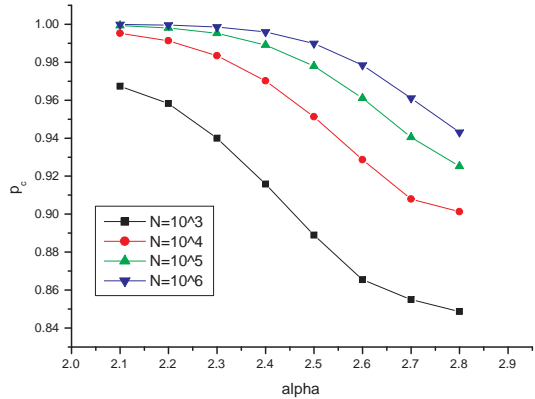,width=7cm}}
\vspace*{8pt} \caption{$p_{c}$ for different $N$ versus $\alpha$
with $m=1$.}
\end{figure}
From Fig. 4, we can see that with the increase of $\alpha$,
$p_{c}$ decreases, while the threshold value $p_{c}$ becomes
bigger with the increase of network scale $N$, which are
consistent with the results shown in Fig.1.

\begin{figure}
\centerline{\psfig{file=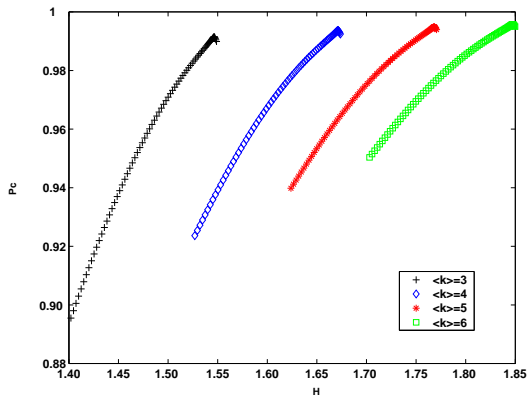,width=7cm}}
\vspace*{8pt}\caption{The relationship between $p_{c}$ and $H$ for
different average connectivity $<k>$. For all curves,
$\alpha\in{(2, 3)}$, $N\in{[10^4, 9*10^4]}$.}
\end{figure}
To get an intuitive understanding of the relationship between the
two measures, we show the results of the threshold value $p_{c}$
and $H$ for different average connectivity $<k>$ in Fig. 5. It can
be seen that $p_{c}$ and $H$ vary consistently, that is, $p_{c}$
increases with $H$ for different average degree $<k>$. Also, with
the increase of $<k>$, both $p_{c}$ and $H$ get improved.

\section{Conclusions}
In real-world networks one is often interested in the question of
how to design a robust network while keeping the cost constant. It
is well known that scale-free networks are resilient to random
failures for their heterogeneous links distributions.

In this paper, different from percolation theory to evaluate the
threshold value $p_{c}$, we use the entropy of the degree
distribution to study the resilience of scale-free networks to
random failures. Heterogeneity is a simple but essential character
of scale-free networks and it is in direct relationship with the
network's resilience to random failures. This character motivates
us to use the entropy of the degree distribution, which provides
an average measure of heterogeneity to study the effect of random
failures on scale-free networks. By optimizing the entropy of the
power-law degree distribution, we get the optimal value of entropy
$H$, the optimal solutions of $\alpha$ and $m$ for different
$<k>$. The results indicate that when the network scale $N$ is
given, the optimal solution $m$ is constant for a certain range of
$<k>$, in which range the scaling exponent $\alpha$ decreases. So
a higher entropy is obtained. Of course, large average
connectivity brings high cost. We then examine the relationship
between threshold value $p_{c}$ and $N$ for different $\alpha$.
The results are also consistent with that of the entropy of the
degree distribution. We also show the relationship of the two
measures$-$$p_{c}$ and $H$. The results indicate that $p_{c}$ and
$H$ vary consistently, that is, $p_{c}$ increases with $H$ for a
certain average degree $<k>$. Also, both $p_{c}$ and $H$ get
improved with the increase of the average degree $<k>$.

We just try an alternative point of view to analyze the robustness
of the network from its heterogeneity. The results indicate that
the entropy of the degree distribution is an effective measure of
scale-free network's resilience to random failures. Deep
discussions on the resilience of scale-free networks to targeted
attacks will be made in the future.

\section*{Acknowledgements}
We are grateful to the anonymous referees for their insightful and
helpful comments. The work was supported by NSFC(90103033).


\end{document}